\documentclass[9pt]{book}

\usepackage[dvips]{graphicx,color}
\usepackage{makeidx,universe}

\makeauthorindex

\BookTitle{The proceedings of the Physics of Accreting Compact Binaries}

\CopyRight{\copyright 2010 by Universal Academy Press, Inc.}

\begin{document} 

\pagenumbering{arabic}

\chapter{%
Dinner Speech: Follow-up in the age of surveys}

\author{\raggedright \baselineskip=10pt%
{\bf Paul J. Groot,$^{1}$}\\ 
{\small \it %
(1) Department of Astrophysics/IMAPP, Radboud University Nijmegen,
  P.O. Box 9010, 6500 GL Nijmegen, The Netherlands
}
}


\AuthorContents{Paul Groot} 

\AuthorIndex{Groot}{P.} 

     \baselineskip=10pt
     \parindent=10pt



\section{Introduction} 

\medskip
Many dinner speeches are not prepared but still given, so this is
probably one of the few that was prepared but never given. The reason
for this is the fact that the conference dinner was held on the roof
of the Kyoto Holiday Inn, where the `beer garden' provided a wonderful
ambience, but where any attempt to words of wisdom would have been immediately
lost on the wind. Committing them to paper is my effort to make them
longer lasting. 

\medskip
First of all I would like to thank the organisers of this meeting, and
in particular Daisaku-san for bringing together such a diverse and
broad community on the topic of accretion physics. Accretion is one of
those `universal' processes that is still poorly understood, but at
the same time encountered in a large variety of astronomical
settings. From the compact binaries discussed here to young stellar
objects, supernovae fall back disks and gamma ray bursts and the
large, massive disks found in active galactic nuclei. 

\medskip
Looking around I see that it is a very active and young community. If
not always young of age then at least of heart! This is a very good
sign, since many challenges still lie before us, and over the next few
years the possibilities in our field will grow as rapidly as the
number of known objects. The field of accreting compact objects is
reinventing itself. After a flurry of activity in the 1980s and 1990s
there appeared to be a slump in activity in the early 2000s. A change
is now occuring from an `object-based' view to a `population-based'
view. No longer are our conferences dominated by talks on single, pet
objects, but more and more the view is widened to address the question
`What does the big picture look like?'. This is a very good and
healthy progression. The field of Cataclysmic Variables (CVs) in
particular was often stigmatized as splitting itself up into yet
another subfield as soon as two new systems were found. Although
questions related to the physics and structure of the myriad of
subpopulations have not always been answered, more and more often a
helicopter view is taken that tries to encompass the populations of
compact binaries as a whole.

\medskip
How do the different populations of compact
binaries (CVs, symbiotics, novae, AM CVns, LMXBs, AMXPs) fit together?
Can we understand underlying processes such as the physics of
the common-envelope by comparing the sizes and characteristics of
different populations with each other? For that matter: do the
properties of a single population already give answers? What
is the relation of compact binaries to other fields in astronomy?
Examples that stand out in this respect are of course the progenitors
of supernovae Type Ia, but also the populations of gravitational wave
emitters and the possibility of new or surviving exoplanets around
white dwarfs. Such a branching out is essential for the continued
health of the field and the anchoring of compact binaries within the
larger community of astrophysics. 

\medskip
During the conference we have already seen very nice examples of this
population view. The almost Sisyphus-like effort of Paula Szkody and
collaborators on uncovering the population of CVs in the Sloan Digital
Sky Survey (SDSS) stands out in this respect. This is a prime example of how
a population based view can answer long-standing questions. The
orbital period distribution of the Sloan CVs as compiled by Boris
G\"ansicke and collaborators clearly shows the existence of the
long-awaited and expected orbital period spike. It also shows the
extreme care one has to take in taking observational selection effects
properly into account: a point already shown quite nicely by the work
of Reta Pretorius and Christian Knigge on the pre-SDSS population of
CVs. 

\section{Big Surveys: the future is near}

\medskip
The SDSS is, however, only the beginning. The near future will see
enormously large synoptic and static surveys that will allow us to
significantly increase the known populations of compact binaries. Not
only of CVs, but also of other, even more rare or relatively
less-studied populations such as AM CVns, pre-CVs, sdBs, LMXBs,
detached binary white dwarfs, AXMPs, novae, etc. Even more importantly, the
systematic design and execution of most of these surveys allows us to
construct observationally well-understood samples of compact binaries
where it is possible to properly model the selection effects and
correct for them. This is absolutely essential for a proper comparison
of observational results with those from e.g. population synthesis
modeling. Any of us designing and/or performing these large surveys
should therefore aim to execute them as systematically and
homogeneously as possible and to properly quantify the selection
effects that went into the design. 

\medskip
Examples of these new and upcoming large surveys that will be very
relevant for our field are the European Galactic Plane Surveys (EGAPS:
IPHAS, UVEX and VPHAS+), Gaia, ATLAS (a Southern Sloan-like survey
with the VST), the Galactic Bulge Survey and the large
synoptic surveys: the Palomar Transient Survey, Pan-Starrs, SkyMapper,
the VVV survey, and also of course Gaia and in the longer run the
LSST. In the radio domain LOFAR will start operation in 2011,
including the Transient Key Project, and proposals for variability
surveys on the SKA-precursors ASKAP (Australia) and MeerKAT (South
Africa) have been submitted. The work presented during this conference
by Brian Warner and Patrick Warner on the Catalina Real Time Transient
(CRTS) Survey and by Martin Still on the Kepler data shows just a
glimpse of what is ahead in these synoptic surveys.

\section{Necessary preparation}

\medskip
As a community we will have to be prepared for these surveys. Not only
by designing efficient algorithms to mine these large databases for
`our' preferred populations, or by having the appropriate theoretical
models to interpret the results, but most importantly to provide
enough resources to execute the task of the photometric and
spectroscopic follow-up on these systems. Many of the targets provided
by the large surveys will require follow-up that is not a part of the
surveys themselves. To characterize systems in terms of masses,
orbital periods, mass-transfer rates, secular evolution and chemical
composition will require vast photometric and spectroscopic
follow-up. We need to be able not only to bring these resources to the
field, but should also try to organize ourselves in such a way that we
can do this follow-up in an efficient and homogeneous way. 

\medskip
Luckily we do, collectively, have the means to do this. We are a world
community. I count participants to this conference from 22 different
countries on 5 different continents. The only continent that is
missing is Australia, but luckily we have Martin Still from outer
space to make up for this. This large and widely-spread community
offers us an enormous opportunity to tackle the challenge before
us. The world will see only 2 or 3 Extremely Large Telescopes, but
there are still many smaller telescopes in the 1-4m class that are
perfectly suited for a large part of the required follow-up. In the
western world the 2-4m class telescopes are under severe funding
pressure, but I am delighted to see that in many other countries new
telescopes in this category are coming into operation. Here we should
tackle the challenge of systematic follow-up by trying to
standardize our follow-up observations as much as possible. We should
be like Japanese vending machines: available on every corner with a
standard output, but they work like a charm! We should try to set-up a
cheap, efficient network of `follow-up vending machines' that deliver
a homogeneous set of data that can easily be intercompared. 

\section{Multi-band photometry}

\medskip
My proposal is {\sl not} to build a new network of telescopes. These
we already have to a large extent. The proposal is to standardize our
follow-up methods and instruments. It strikes me again and again how
often we `waste' astronomical information because our instruments are
not designed to take full advantage of what modern technology can
offer. Also during this conference we have seen beautiful photometric follow-up
studies where on a particular object, for instance, 3000 frames have been
taken, of which 2800 in, say, $R$, 200 in $V$ and 2 in $I$. This is
certainly not unique to our field. I see the same happening in the
gamma-ray burst and supernovae follow-up, although the last is
slightly better since the MCSL method for calibrating Supernovae Type
Ia requires multi-band data. 

\medskip
Modern technology, however, should allow us to build a simple
multi-band photometer that covers, e.g. the SDSS bands simultaneously
and allows for a 1 second read-out of the CCDs. This is the type of
instrument that we should have as a standard 'photometer' on as many
of the 1-4m class telescopes in the world. Only when we these
instruments will it be possible to come to a homogeneous approach to
the photometric follow-up of the thousands of targets that will be
provided by the next generation of large scale surveys.

\section{Closing}

\medskip
In closing I would like to say again how much I am encouraged by the
diversity of researchers present here. It has been a week of high
spirits, and enthousiasm, and in this, I would like to say that we, as
a community, should also take our Japanese hosts as an example. I have
rarely been in a country where the cultural and language barriers
appeared to be so large, but where the crossing of these barriers is
undertaken with such an enthousiasm and such friendliness. 
 
\bigskip
{\sl Domo arigato} and {\sl itadakimasu}

\end{document}